# Differences between Two Maximal Principal Strain Rate Calculation Schemes in Traumatic Brain Analysis with in-vivo and in-silico Datasets

Xianghao Zhan[2,#], Zhou Zhou[2,3,#], Yuzhe Liu[1,2,*], Nicholas J. Cecchi[2], Marzieh Hajiahamemar[4], Michael M. Zeineh[5], Gerald A. Grant[5,6,7], David Camarillo[2,6,8]

(1) School of Biological Science and Medical Engineering, Beihang University, Beijing, 10019, China
(2) Department of Bioengineering, Stanford University, CA, 94305, USA
(3) Division of Neuronic Engineering, KTH Royal Institute of Technology, Stockholm, SE-100 44, Sweden
(4) Department of Biomedical Engineering & Chemical Engineering, University of Texas at San Antonio, San Antonio, TX, 78249, USA
(5) Department of Radiology, Stanford University, CA, 94305, USA
(6) Department of Neurology, Stanford University, Stanford, CA, 94305, USA.
(7) Department of Neurosurgery, Duke University, Durham, NC, 27710, USA
(8) Department of Mechanical Engineering, Stanford University, Stanford, CA, 94305, USA.

# These authors contributed equally to this manuscript.
* Corresponding Author: Yuzhe Liu (yuzheliu@buaa.edu.cn, yuzheliu@stanford.edu)

## Abstract

Brain deformation caused by a head impact leads to traumatic brain injury (TBI). The maximum principal strain (MPS) was used to measure the extent of brain deformation and predict injury, and the recent evidence has indicated that incorporating the maximum principal strain rate (MPSR) and the product of MPS and MPSR, denoted as MPS×SR, enhances the accuracy of TBI prediction. However, ambiguities have arisen about the calculation of MPSR. Two schemes have been utilized: one ($MPSR_1$) is to use the time derivative of MPS, and another ($MPSR_2$) is to use the first eigenvalue of the strain rate tensor. Both $MPSR_1$ and $MPSR_2$ have been applied in previous studies to predict TBI. To quantify the discrepancies between these two methodologies, we conducted a comparison of these two MPSR methodologies across nine in-vivo and in-silico head impact datasets and found that $95MPSR_1$ was 5.87% larger than $95MPSR_2$, and $95MPS \times SR_1$ was 2.55% larger than $95MPS \times SR_2$. Across every element in all head impacts, $MPSR_1$ was 8.28% smaller than $MPSR_2$, and $MPS \times SR_1$ was 8.11% smaller than $MPS \times SR_2$. Furthermore, logistic regression models were trained to predict TBI based on the MPSR (or MPS×SR), and no significant difference was observed in the predictability across different variables. The consequence of misuse of MPSR and MPS×SR thresholds (i.e. compare threshold of $95MPSR_1$ with value from $95MPSR_2$ to determine if the impact is injurious) was investigated, and the resulting false rates were found to be around 1%. The evidence suggested that these two methodologies were not significantly different in detecting TBI.

## Introduction

In traumatic brain injury (TBI), brain deformation can be calculated using validated finite element (FE) human head models, with impact kinematics measured by wearable sensors as input (Zhan et al., 2023; Zhan et al., 2022a; Zhan et al., 2022b). Recent evidence has suggested that severe brain deformation is associated with neuropathology, including axonal injury (Cullen et al., 2016; Hajiaghamemar and Margulies, 2021; Hajiaghamemar et al., 2020a; Hajiaghamemar et al., 2020b; Miller et al., 2022) and blood-brain barrier disruption (O'Keeffe et al., 2020). Therefore, researchers have dedicated their efforts to predicting TBI by identifying mechanical variables that are more relevant to injury prediction (Ghazi et al., 2021; Giordano et al., 2014; Hernandez et al., 2015; Kleiven, 2007; Patton et al., 2013; Wu et al., 2019; Wu et al., 2022a; Wu et al., 2022b). Deformation is typically described as a strain tensor with six degrees of freedom, and the first eigenvalue of the strain sensor, which is also known as maximal principal strain, has been used to indicate brain injury risk. In most studies, the peak maximal principal strain over time has been used to represent the most severe deformation that the local brain tissue has experienced, which is denoted as MPS and the 95$^{th}$ percentile of peak MPS over the whole brain (denoted as 95MPS) is typically used to gauge the overall severity of brain deformation and mitigate the impact of potential numerical error on the outcomes (compared with the maximum MPS) (Kleiven, 2007; Li et al., 2021; Miller et al., 2022; Patton et al., 2013; Zhan et al., 2023).

Recent findings from human and animal studies have used maximal principal strain rate (MPSR, similar to MPS, MPSR indicates the value over time) - a measure of how quickly brain deformation changes as well as the product of maximum principal strain and its rate (MPS×SR, which is also the peak value over time), which both hold greater predictive value for TBI than MPS (Hajiaghamemar and Margulies, 2021; Hajiaghamemar et al., 2020a; Hajiaghamemar et al., 2020b; O'Keeffe et al., 2020; Wu et al., 2022b). However, ambiguity arises about how MPSR is calculated (Zhou et al., 2023). Two computational schemes are commonly implemented: scheme 1 is to directly differentiate maximal principal strain against time and find the maximum over time (MPSR$_1$), and scheme 2 is to use the first eigenvalue of the strain rate tensor, which is the symmetric part of the velocity gradient, and find its maximum over time (MPSR$_2$). The difference between the calculation of MPSR$_1$ and MPSR$_2$ is demonstrated in **Fig.1** (Dill, 2006). Along the same lines, there are two corresponding schemes to calculate MPS×SR, which are denoted as MPS×SR$_1$ and MPS×SR$_2$. Additionally, the 95$^{th}$ percentile highest values observed across the entire brain (denoted as 95MPS, 95MPSR, 95MPS×SR) were used to represent the overall severity of deformation of the whole brain.

Although MPSR$_1$ and MPSR$_2$ exhibit distinct calculation methods, both have been used to assess brain injury in research. The variables have been often denoted simply as "MPSR" or "MPS×SR" without differentiation (**Table 1**). Recognizing the importance of this distinction and the need to quantify it, we previously performed theoretical clarification of two strain rate computational schemes and substantiated the difference by cadaveric experiments and only one concussive impact (Zhou et al., 2023). Here, we extended our effort by calculating the differences between the two schemes based on nine head impact datasets (*n*= 3873) with the KTH head model (Kleiven, 2007) and evaluated differences at both the brain-element level and the whole-brain based on MPSR and MPS×SR. Furthermore, we tested the goodness of fit and the predictability of MPS, as well as the two schemes of MPSR and MPS×SR. Additionally, we estimated how many cases would be misclassified if the calculation schemes were misused when determining injurious cases with thresholds.

**Method**

To quantify the differences in different MPSR calculation schemes, we used nine published datasets of head impact kinematics in different types of head impact. For the head kinematics measured by instrumented mouthguard (Camarillo et al., 2013; Liu et al., 2020), we included two datasets for college football (CF1, n=118 (Liu et al., 2021) and CF2, n=184 (Laksari et al., 2020)), one dataset for high school football (HSF, n=602 (Cecchi et al., 2021)), two datasets for mixed martial arts games (MMA-G, n=79 (O'Keeffe et al., 2020)) and practices (MMA-P, n=384 (Tiernan et al., 2020)). For the head kinematics measured by reconstructed head impacts in labs, we included a dataset of National Football League head impacts (NFL, n=53 (Sanchez et al., 2019)), in which injury information was recorded, and a dataset from the impact tests in National Highway Traffic Safety Administration (NHTSA n=48 (Zhan et al., 2021a)). We also included a dataset collected from numerically reconstructed accidents in the National Association for Stock Car Auto Racing (NASCAR n=275(Somers et al., 2011)). Furthermore, we included a head impact dataset generated by FE simulations of linear impactor tests, and this dataset covered a large range of impact speeds and directions (HM, short for head model, n=2130 (Zhan et al., 2021b)).

Brain deformations were calculated by the KTH head model (Kleiven, 2007), which is a validated FE model including 4124 elements representing the brain. The simulation protocol was similar to the previous study (Zhan et al., 2024). For every element, MPSR$_1$ was calculated as the peak of the time derivative (five-point stencil with a time interval of 1 ms) of MPS, which was exported as the 1st-principal component of the Green strain tensor, and MPSR$_2$ was exported as the peak of 1st-principal component of the strain rate tensor. Then, MPS×SR was calculated as the peak of the maximum principal strain multiplied by the maximal principal strain rate over every time point (**Fig.1**).

The influences of MPSR computational schemes were compared in three ways: (a) element levels: $MPSR_1$ ($MPS \times SR_1$) are compared with $MPSR_2$ ($MPS \times SR_2$) at every brain element; (b) on the whole-brain level: $95MPSR_1$ ($95MPS \times SR_1$) were compared with $95MPSR_2$ ($95MPS \times SR_2$) for every impact. The differences were plotted in the Bland-Altman plot. For each dataset, the values given by the two schemes were compared by pairwise t-tests (normality confirmed beforehand), and the differences in the averages were calculated. (c) Based on the NFL dataset, in which the TBI diagnosis information was recorded for each case (22 injurious head impacts and 31 non-injurious head impacts), the strain rate variables ($95MPSR_1$, $95MPSR_2$, $95MPS \times SR_1$, $95MPS \times SR_2$) were used to fit logistic regression models, and the deviances for every model were calculated to show the goodness of fit of the model. Then, the NFL dataset was randomly partitioned into training (60%) and test (40%) datasets for 40 rounds. In each round, the logistic regression model was developed on the training dataset, and the thresholds corresponding to 50% risk were used to classify the injurious and non-injurious head impacts in the test dataset. The accuracy, precision, recall, and F1 of different variables were compared. Additionally, we investigated how many cases are misclassified if the variable was compared with the threshold from different schemes. There are four different scenarios of misusing schemes: SN1: $95MPSR_2$ is compared with $95MPSR_1$ threshold; SN2: $95MPSR_1$ is compared with $95MPSR_2$ threshold; SN3: $95MPS \times SR_2$ is compared with $95MPS \times SR_1$ threshold; SN4: $95MPS \times SR_1$ is compared with $95MPS \times SR_2$ threshold. The thresholds were for 50% risk according to the logistic regression model fitted on the NFL dataset. The rates of misclassification in every dataset were calculated and compared.

**Results**
*Peak MPSR and MPS×SR at every element*
**Table 2** lists the difference between the two schemes at the element level, and the Bland-Altman plots of the comparisons are given in **Fig.S1,2**. Similar patterns could be observed in the distribution of differences: the differences were small in most elements but got large in a few elements. In more than 60% of elements, $MPSR_1$ was lower than $MPSR_2$, and the average of $MPSR_1$ across all elements was also lower than the average of $MPSR_2$. Similar results could be found in the comparison between $MPS \times SR_1$ and $MPS \times SR_2$ (**Fig.S2**), except for CF2 and HM datasets. On average across every element in every impact, the $MPSR_1$ was smaller than $MPSR_2$ (ME: -1.68 s$^{-1}$, MPE: -12.73%), and the $MPS \times SR_1$ was smaller than $MPS \times SR_2$ (ME: -0.01 s$^{-1}$, MPE: -11.95%).

*95$^{th}$ Percentile Peak MPSR and MPS×SR for every impact*
**Table 3** lists the difference between the two schemes in 95MPSR, and the Bland-Altman plots of the comparisons are given in **Fig.S3,4**. The average $95MPSR_1$ was lower than the average $95MPSR_2$ in all datasets except CF2, NFL, and HM. Furthermore, $95MPSR_1$ and $95MPSR_2$ were significantly different in most cases ($p<0.05$), except for datasets CF2 and NASCAR. When comparing 95MPS×SR, the difference ($95MPS \times SR_1 – 95MPS \times SR_2$) was positive in most datasets except for MMA-P (**Fig.5**). $95MPS \times SR_1$ and $95MPS \times SR_2$ were significantly different ($p<0.05$), except for in datasets CF1, MMA-G and HSF. On average across every impact, the $95MPSR_1$ was close to $95MPSR_2$, and the $95MPS \times SR_1$ was 4.85% larger than $95MPS \times SR_2$.

*Logistic Regression Based on NFL Dataset*
**Table 4** lists the deviances of logistic regression models using different strain rate parameters. The regression models are shown in **Fig.S5**. Slightly lower deviance was found in scheme 2, and the

95MPS×SR$_2$ had the lowest deviance. Then, the NFL dataset was partitioned into training and test datasets to evaluate the predictability of different strain rate variables. 1-way ANOVA was performed in comparison, and no significant difference was observed in accuracy, precision, recall, and F1 (averages were listed in **Table 4** and box plots were in **Fig.S6**).

*False Classification by Using Different Schemes*
When comparing the strain rate variables with the thresholds given by different schemes, a small rate of misclassified cases was found: the false rate for 95MPSR was 1.34% (averaged by impacts) and 1.07% (by datasets), and the false rate for 95MPSR×SR was 1.24% (by impacts) and 1.11% (by datasets). High false rates were found in the NHTSA and NFL datasets. The box plots are given in **Fig.S7**.

**Discussion:**
In this study, we compared two schemes to calculate MPSR, which has been widely used as a mechanical variable for TBI prediction. Scheme 1 considers the "maximum principal strain" as a scalar value, making MPSR the rate of change of MPS. Mathematically, it is computed by directly differentiating MPS over time. Scheme 2 views strain rate as a tensor, with MPSR representing the maximal principal value of the strain rate tensor. The utilization of strain rate and the multiplication of strain rate and strain were initially introduced in TBI studies (Newman et al., 1999) based on theoretical analysis (Viano and Lau, 1988). Subsequently, the authors (Newman et al., 1999) started to use scheme 1 in their studies. Scheme 2, conversely, is more widely used in solid mechanics, and the direct output of FE software "1st principal strain rate" is calculated in scheme 2, so it has been used by other groups (**Table 1**). It is important to note that schemes 1 and 2 give the same results if the principal axes of strain tensor and strain rate tensor remain constant, such as single-axis tension. When the principal axes of strain rate tensor change rapidly, large difference may be observed between schemes 1 and 2. Furthermore, because MPSR$_2$ is the maximal of principal values, it is always non-negative. In contrast, MPSR$_1$ can be negative when MPS decreases. From the various aspects mentioned above, MPSR$_1$ and MPSR$_2$ are distinctive variables with different physical meanings. Nevertheless, given that the underlying mechanical mechanisms responsible for TBI remain largely unclear, determining which of these variables better captures the "injury signal" remains to be a challenge. As both schemes were adopted in previous studies (**Table 1**), researchers should be aware of the difference between them when interpreting and analyzing the published results.

In the logistic regression of TBI injury and strain rate variables, slightly different deviances were observed. 95MPSR has a better fit to injury than 95MPS×SR, and both of them have a better fit than 95MPS. In both 95MPSR and 95MPS×SR, scheme 2 exhibits better fits than scheme 1. However, the differences in deviance did not result in significant differences in predictability. Furthermore, the false rate caused by the misuse of thresholds was estimated as ~1%, which is acceptable compared with the uncertainty of TBI prediction. However, the differences in schemes should be aware when the studies focus on the spatial distribution or the impact process.

In this study, one limitation is that we did not include axonal strain rate (tract-oriented strain rate) in comparison because we do not have access to an FE head model with axonal fibers. Another limitation is that we only compared the fitness of logistic regression models instead of testing the predictability because of the small size of the NFL dataset.

**Declarations**

**Competing interests**  The authors declare that they have no conflicts of interest.

**Acknowledgment**

This work was supported by the National Natural Science Foundation of China (NSFC 12302414), Young Elite Scientists Sponsorship Program by The Chinese Society of Theoretical and Applied Mechanics (CSTAM 2023-XSC-HW3), the Fundamental Research Funds for the Central Universities Program (YWF-23-Q-1029), the Pac-12 Conference's Student-Athlete Health and Well-Being Initiative, the National Institutes of Health (R24NS098518), Taube Stanford Children's Concussion Initiative and Stanford Department of Bioengineering.

**Figures and Tables**

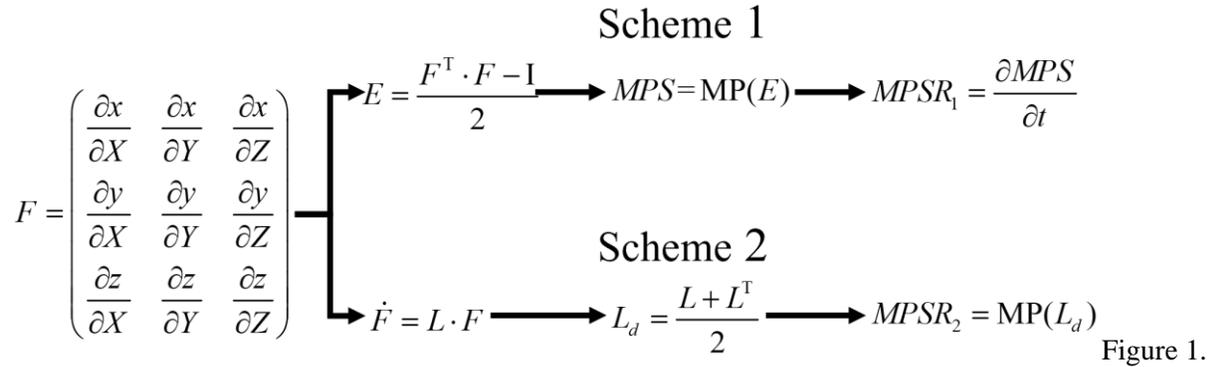

Figure 1. The difference between the calculation of MPSR$_1$ and MPSR$_2$. $F$ is the deformation gradient and $\dot{F}$ is the time derivative of the deformation gradient; superscript T indicates transpose. ($x$, $y$, $z$) are the deformed configurations, ($X$, $Y$, $Z$) are the original configuration, $E$ is the Green-Lagrange strain, MP( ) is the function to calculate the largest eigenvalue of the matrix, $L$ is the velocity gradient, and $L_d$ is the strain rate tensor defined implicitly by $\dot{F} = L \cdot F$ (Dill, 2006).

Table 1. Studies calculating MPSR and MPS×SR in different schemes.

| Reference | MPSR | Head Impacts | FE model |
|---|---|---|---|
| (Hajiaghamemar and Margulies, 2021) | Scheme 1 | Preclinical | pig brain model |
| (Hajiaghamemar et al., 2020b) | | Preclinical | pig head model |
| (Wu et al., 2022b) | | NFL (reconstruction); Naval sled tests (on-field); Preclinical | GHBMC, pig, macaque, baboon brain models |
| (Wu et al., 2021) | | NFL (reconstruction); Naval sled tests (on-field); Preclinical | GHBMC, pig, macaque, baboon brain models |
| (Miller et al., 2022) | | American Football (HITS) | Atlas-based brain model |
| (Miller et al., 2021) | | American Football (HITS) | Atlas-based brain model |
| (Kuo et al., 2018) | | PMHS (sensor, MiG) post-mortem human surrogate | WHIM |
| (King et al., 2003) | | NFL (reconstruction) | WSUBIM |
| (Hernandez et al., 2019) | | American Football (MiG) | KTH |
| (Zhang et al., 2004) | | NFL (reconstruction) | WSUBIM |
| (Sahoo et al., 2016) | | Traffic Accidents and American Football (reconstruction) | ULP |
| (O'Keeffe et al., 2020) | Scheme 2 | Mixed Martial Arts (MiG) | KTH |
| (Zhou et al., 2022) | | American Football (MiG) | KTH-detailed model |
| (Zhan et al., 2022a) | | American Football, Mixed Martial Arts (MiG); Headform Impact, NASCAR (Simulation) | KTH |
| (Liu et al., 2021) | | American Football (MiG) | KTH |
| (Ji et al., 2022) | | NFL (reconstruction) | WHIM |
| (McAllister et al., 2012) | unknown | American Football, Hockey (HITS) | Dartmouth Model |
| (Beckwith et al., 2018) | | American Football (HITS) | WHIM |
| (Wu et al., 2022a) | | American Football (MiG, HITS) | WHIM |

Table 2. Statistics of differences between $MPSR_1$ and $MPSR_2$ ($MPSR_1$ - $MPSR_2$) and the difference between $MPS \times SR_1$ and $MPS \times SR_2$ ($MPS \times SR_1$ - $MPS \times SR_2$) for every element in every head impact. (ME: mean error; MAE: mean absolute error; RSME: root square mean error; MAPE: mean absolute peak error). The Bland-Altman plots of $MPSR_1$ and $MPSR_2$ were given in **Fig.S1**, and the Bland-Altman plots of $MPS \times SR_1$ and $MPS \times SR_2$ were given in **Fig.S2**.

| | $MPSR_1$ - $MPSR_2$ | | | | | | |
|---|---|---|---|---|---|---|---|
| Dataset | ME (1/s) | MAE (1/s) | RSME (1/s) | MPE (%) | MAPE (%) | LoA Upper (1/s) | LoA Lower (1/s) |
| CF1 | -2.49 | 2.91 | 5.50 | -19.38 | 21.17 | 7.12 | -12.10 |
| CF2 | -1.28 | 2.31 | 5.19 | -16.97 | 19.52 | 8.58 | -11.14 |
| HSF | -3.28 | 5.32 | 109.56 | -16.82 | 21.02 | 211.37 | -217.92 |
| MMA-G | -6.25 | 8.02 | 16.08 | -22.40 | 25.02 | 22.79 | -35.28 |
| MMA-P | -2.57 | 2.81 | 4.47 | -22.51 | 23.79 | 4.59 | -9.74 |
| NFL | -1.88 | 4.11 | 6.03 | -12.59 | 19.99 | 9.35 | -13.11 |
| NHTSA | -5.73 | 8.37 | 27.51 | -14.14 | 21.39 | 47.02 | -58.47 |
| NASCAR | -2.31 | 3.38 | 5.93 | -15.89 | 19.13 | 8.39 | -13.00 |
| HM | -0.72 | 2.71 | 5.14 | -8.28 | 14.92 | 9.26 | -10.70 |
| Average by Impacts | -1.68 | 3.35 | 43.62 | -12.73 | 17.81 | 83.74 | -87.11 |
| Average by Datasets | -2.82 | 4.33 | 22.90 | -16.17 | 20.38 | 41.22 | -46.86 |
| | $MPS \times SR_1$ - $MPS \times SR_2$ | | | | | | |
| Dataset | ME (1/s) | MAE (1/s) | RSME (1/s) | MPE (%) | MAPE (%) | LoA Upper (1/s) | LoA Lower (1/s) |
| CF1 | -0.06 | 0.17 | 0.67 | -17.56 | 21.14 | 1.24 | -1.36 |
| CF2 | 0.06 | 0.25 | 1.43 | -16.32 | 20.95 | 2.86 | -2.74 |
| HSF | -0.21 | 0.90 | 40.15 | -14.95 | 22.19 | 78.47 | -78.89 |
| MMA-G | -0.29 | 0.89 | 3.88 | -20.57 | 25.12 | 7.29 | -7.87 |
| MMA-P | -0.07 | 0.10 | 0.24 | -21.14 | 23.61 | 0.39 | -0.53 |
| NFL | -0.02 | 0.43 | 0.93 | -13.16 | 22.63 | 1.80 | -1.85 |
| NHTSA | -0.05 | 0.83 | 2.93 | -8.75 | 22.54 | 5.69 | -5.78 |
| NASCAR | -0.05 | 0.31 | 1.14 | -14.83 | 20.72 | 2.18 | -2.29 |
| HM | 0.07 | 0.36 | 0.98 | -8.11 | 18.15 | 1.98 | -1.84 |
| Average by Impacts | -0.01 | 0.42 | 15.86 | -11.95 | 19.99 | 31.08 | -31.10 |
| Average by Datasets | -0.06 | 0.47 | 6.82 | -14.73 | 21.70 | 13.30 | -13.43 |

Table 3. Statistics of differences between 95MPSR$_1$ and 95MPSR$_2$ (95MPSR$_1$ - 95MPSR$_2$) and differences between 95MPS×SR$_1$ and 95MPS×SR$_2$ (95MPS×SR$_1$ - 95MPS×SR$_2$) for every head impact. (ME: mean error; MAE: mean absolute error; RSME: root square mean error; MAPE: mean absolute peak error). The Bland-Altman plots of 95MPSR$_1$ and 95MPSR$_2$ were given in **Fig.S3**, and the Bland-Altman plots of 95MPS×SR$_1$ and 95MPS×SR$_2$ were given in **Fig.S4**.

| | 95MPSR$_1$ - 95MPSR$_2$ | | | | | | |
|---|---|---|---|---|---|---|---|
| Dataset | ME (1/s) | MAE (1/s) | RSME (1/s) | MPE (%) | MAPE (%) | LoA Upper (1/s) | LoA Lower (1/s) |
| CF1 | -2.20 | 3.35 | 5.76 | -9.69 | 12.54 | 8.28 | -12.69 |
| CF2 | 0.61 | 2.67 | 7.36 | -5.34 | 9.90 | 15.02 | -13.80 |
| HSF | -3.94 | 9.82 | 49.01 | -6.01 | 13.80 | 91.88 | -99.76 |
| MMA-G | -6.07 | 9.35 | 17.44 | -11.99 | 14.49 | 26.19 | -38.33 |
| MMA-P | -2.70 | 3.01 | 4.97 | -12.93 | 13.93 | 5.50 | -10.90 |
| NFL | 2.01 | 3.38 | 4.37 | 3.05 | 9.53 | 9.69 | -5.68 |
| NHTSA | -10.19 | 18.30 | 49.34 | 0.51 | 16.20 | 85.43 | -105.81 |
| NASCAR | -0.06 | 2.37 | 3.46 | -3.62 | 8.17 | 6.73 | -6.84 |
| HM | 2.76 | 3.45 | 6.19 | 5.87 | 9.44 | 13.63 | -8.11 |
| Average by Impacts | 0.37 | 4.58 | 20.93 | 0.01 | 10.78 | 41.39 | -40.64 |
| Average by Datasets | -1.94 | 6.03 | 16.88 | -4.01 | 11.88 | 30.37 | -34.26 |
| | 95MPS×SR$_1$ - 95MPS×SR$_2$ | | | | | | |
| Dataset | ME (1/s) | MAE (1/s) | RSME (1/s) | MPE (%) | MAPE (%) | LoA Upper (1/s) | LoA Lower (1/s) |
| CF1 | 0.09 | 0.28 | 0.93 | -5.16 | 11.63 | 1.90 | -1.73 |
| CF2 | 0.57 | 0.65 | 2.90 | -1.85 | 11.43 | 6.15 | -5.01 |
| HSF | 0.42 | 2.10 | 9.01 | -0.58 | 15.57 | 18.08 | -17.23 |
| MMA-G | 0.40 | 1.22 | 3.93 | -5.53 | 11.38 | 8.10 | -7.30 |
| MMA-P | -0.05 | 0.12 | 0.28 | -8.65 | 11.23 | 0.49 | -0.59 |
| NFL | 0.67 | 0.78 | 1.25 | 6.97 | 13.66 | 2.76 | -1.41 |
| NHTSA | 1.15 | 1.92 | 3.07 | 13.75 | 21.14 | 6.79 | -4.50 |
| NASCAR | 0.32 | 0.45 | 1.11 | -0.66 | 10.55 | 2.42 | -1.77 |
| HM | 0.83 | 0.87 | 1.67 | 10.81 | 14.62 | 3.68 | -2.02 |
| Average by Impacts | 0.60 | 0.95 | 3.89 | 4.85 | 13.90 | 8.13 | -6.93 |
| Average by Datasets | 0.50 | 0.93 | 2.80 | 1.40 | 13.51 | 5.85 | -4.85 |

Table 4. The deviances of the logistic regression model on the NFL dataset, and the mean accuracy, precision, recall, and F1 of the classification using thresholds corresponding to 50% risk in logistic regression models. The logistic regression models fitted on the whole NFL dataset were given in **Fig.S5**, and the box plots of comparison of accuracy, precision, recall, and F1 were given in **Fig.S6**.

|  | Deviance | Mean Accuracy | Mean Precision | Mean Recall | Mean F1 |
|---|---|---|---|---|---|
| 95MPSR$_1$ | 42.0 | 0.83 | 0.82 | 0.76 | 0.78 |
| 95MPSR$_2$ | 41.1 | 0.84 | 0.84 | 0.75 | 0.78 |
| 95MPS×SR$_1$ | 42.1 | 0.81 | 0.81 | 0.69 | 0.73 |
| 95MPS×SR$_2$ | 41.5 | 0.81 | 0.81 | 0.71 | 0.74 |